# Terahertz Nonlinear Superconducting Metamaterials

Caihong Zhang,[1] Biaobing Jin,[2] Jiaguang Han,[1,3] Iwao Kawayama,[1] Hiro Murakami,[1] Jingbo Wu,[2] Lin Kang,[2] Jian Chen,[2] Peiheng Wu,[2] and Masayoshi Tonouchi[1,a]

[1]*Institute of Laser Engineering, Osaka University, 2-6 Yamadaoka, Suita, Osaka 565-0871, Japan*
[2]*Research Institute of Superconductor Electronics, Nanjing University, Nanjing 210093, China*
[3]*Center for Terahertz Waves and College of Precision Instrument and Optoelectronics Engineering, Tianjin University, Tianjin 300072, China*

**Abstract**

We investigate the nonlinear effect of a planar superconducting metamaterial made from niobium nitride (NbN) at terahertz frequencies. As the variation of the incident intense terahertz field alters the intrinsic conductivity in the NbN, a consequent giant amplitude modulation is observed due to the strong nonlinearities. The high sensitivity of the chosen metamaterial even allows observing the nonlinear behaviors at various temperatures, but the resonance modulation induced by the nonlinear effect was distinct from that induced by the heating effect. The presented results illustrate a clever implementation of strongly enhanced nonlinearities and thus may bring nonlinear metamaterials into novel applications.

[a] Electronic mail: tonouchi@ile.osaka-u.ac.jp

A wide variety of intriguing effects, such as negative refraction,[1-4] cloaking,[5] and super-focusing,[6] can be obtained with simple linear metamaterials, where it has been implicitly assumed that the metamaterials interact with electromagnetic waves only through linear processes. However, the notation of metamaterials can actually be extended to the nonlinear regime if the element is formed from or contains nonlinear components whose permittivities are dependent on the intensity of light. A nonlinear metamaterial is no doubt to open the way to the possibility of exploiting nonlinear optical phenomena and further developing new applications. Hence, the nonlinear metamaterials is recently a rapidly developing field of enormous interest ranging from microwaves to optics.

Among them, the terahertz regime is especially attractive for the nonlinear metamaterials research because of the unique permittivity of materials at terahertz frequencies. In particular, the recent development of high-field terahertz sources makes it possible for one to exploit nonlinear effects, and dramatic nonlinearities in conventional matter, such as semiconductors,[7] quantum wells,[8] ferroelectrics,[9] metal oxides,[10] *etc*, have been observed. Indeed, a metamaterial is arguably the ideal environment for nonlinearity because it allows the light to be confined tightly over the scale of subwavelength and thereby the high field intensities permit a greater sensitivity to small changes of materials. As one important example, in this letter, we explore the nonlinear response of a superconducting (SC) metamaterial in the terahertz regime. The proposed metamaterial structure is comprised of an array of subwavelength split-ring resonators (SRRs) made from superconducting niobium nitride (NbN) film on MgO substrate and the nonlinear response lies in the suppression of the superconductivity of NbN by the intense terahertz wave. Increasing or decreasing the incident terahertz field strength, one can creates a sharper "off" or "on" transmission of the chosen SC metamaterial at the resonance frequency. In addition, the nonlinear response is also investigated at various temperatures and it was found that the heating effect affects the spectra of the SC metamaterial more significantly compared to that induced by the nonlinear effect. The presented nonlinear SC metamaterial not only can offer far lower losses and extreme sensitivity to the subtle changes of superconducting state with external stimuli such as thermal tuning,[11-13] electric[14] and magnetic fields,[15] but also allows to dynamically switch metamaterials using light by taking advantage of the strong and fast nonlinear response of lumped elements.[16,20] The nonlinear SC metamaterials thus open up new opportunities for manipulating electromagnetic waves and promise a fascinating prospect on novel device developments and applications.

The essence of the chosen metamaterial is a planar array of superconducting NbN SRRs periodically printed on a 1 mm-thick MgO substrate. Figure 1(a) shows a single SRR unit, here, $g = t = 5$ µm, $w = 10$ µm, and $a = 50$ µm. The SRRs are printed, as shown in Fig. 1(b), on a square lattice of period $p = 60$ µm. The SRRs are made of 50 nm-thick NbN film (superconducting transition temperature $T_c$=15.4 K), where the thin film was firstly deposited on the MgO substrate using RF magnetron sputtering and then was patterned with standard photolithograph and reactive ion etching (RIE) method.

For the nonlinear transmission measurements, high-field terahertz experiments were performed. The intense single-cycle terahertz radiation was generated by optical rectification in a LiNbO$_3$ crystal using tilted pulse-front excitation.[17] The experimental setup consists of a 100-fs Ti: Sapphire regenerative amplifier (Spitfire, Spectra-Physics) operating at 800 nm with a repetition of 1 kHz, and a conventional terahertz detection system based on the electro-optical sampling technique. The terahertz beam was tightly focused onto the cryogenically cooled samples over a temperature range from 4.2 to 300 K. Using a pyroelectric detector (SPI-A-62,

Spectrum Detector Inc.), we estimated the peak terahertz field strength of approximately $E_0$=30 kV/cm. The incident terahertz field strength $E_{in}$ could be varied fast from about 1 to 30 kV/cm by using a pair of terahertz wire grid polarizers driven by an electric motor.

A sharp inductor-capacitor (LC) resonance is excited if the incident terahertz electric field is perpendicular to the gap of the SRR. The amplitude transmission is obtained by $|\tilde{t}(\omega)| = |\vec{E}_S(\omega)/\vec{E}_R(\omega)|$, where $\vec{E}_S(\omega)$ and $\vec{E}_R(\omega)$ are Fourier transformed transmitted electric field of the sample and reference (bare MgO substrate) pulses, respectively. Figure 2(a) shows the measured transmission spectra $|\tilde{t}(\omega)|$ through the samples for various values of $E_{in}$ at 4.5 K. At low incident field strength ($E_0$/16), which is too weak to induce the nonlinear effect,[18] the transmission shows a sharp resonance dip approaching -20 dB located at about 0.45 THz. As the incident terahertz field strength is increased from $E_0$/16 to $E_0$, the resonant transmission dip decreases remarkably from -20 dB to -3 dB without prominent resonance frequency shift. At $E_{in}$ = $E_0$, the LC resonance dip was significantly decreased and even absent. A very substantial change of more than 90% in the resonant transmission has been achieved through increasing the incident field strength, which is attributed to the nonlinearity of the lumped superconductor NbN elements. Such a SC metamaterials therefore allows to simply and fast switch from a low-transmission state to a high-transmission state by adding or removing power from the incident terahertz wave, and vice versa.

To understand the strength-dependent behavior of the SC metamaterial, we also measured the NbN thin film with the same thickness of 50 nm under various $E_{in}$ at 4.5 K. The measured complex conductivity ($\sigma = \sigma_1 - j\sigma_2$) of the SC NbN film is extracted and showed in Fig. 2(c) and (d). The strong terahertz radiation causes breakup of Coop-pairs,[18-19] and thus $\sigma_1$ increases but $\sigma_2$ decreases with increasing $E_{in}$, which fulfill the two-fluid model.[12-13] With the measured complex conductivity data of the NbN thin film, computer simulations of the spectral response of the chosen SC metamaterial were performed using the commercial software CST Microwave Studio. The unit cell shown in Fig. 1(a) was taken into account in the simulations with a periodic boundary condition. Fig. 2(b) shows the simulated spectra $|\tilde{t}(\omega)|$, which reveals good agreement with the experimental data in Fig. 2(a), which further confirms that the modulation of the amplitude transmission is essentially associated with the changes in the conductivity of the SC NbN thin film due to the nonlinearity.

To acquire more nonlinear characteristics of the chosen SC metamaterial, we measured the sample at different temperatures. Figure 3(a) and (b) show the measured transmission spectra for the various terahertz field strengths at 10 K and 13 K, respectively, which are qualitatively similar to that at 4.5 K in Fig. 2(a). The resonant transmission gets a rise as the incident terahertz field strength increases, but the range of variation declines dramatically with the increase of the temperature. Hence, although the similar nonlinear effect could be observed at different temperatures, it is apparent that the LC resonance appears to attenuate as the temperature is increased towards $T_c$ and the nonlinearity thus becomes weaker than that at a lower temperature. The results implied that the heating effect shows stronger influences on the spectra response of the chosen SC metamaterial. Specifically, here we fixed the incident terahertz field strength to the low value $E_0/16$, and measured the transmission at different temperatures varying from 4.5 K to 15.5 K. Figure 3 (c) shows the temperature-dependent transmission of the SC metamaterial, and it is clearly seen that as the temperature increases, the resonance dip decreases obviously and also even disappears around $T_c$. However, unlike those observed in the field strength-dependent measurements, changing the temperature not only decreases the resonance dip, but also leads to notable redshift of the resonance frequency, which is shifted from 0.45 to 0.36 THz when the temperature is increased from 4.5 to 14 K. Because no prominent nonlinear effect would occur at $E_{in}= E_0/16$, the variations here mainly originate from the heating effect.[18] Actually, it is known that the temperature also can give rise to the significant changes of the conductivity of the NbN film due to the heating effect, which in turn causes pronounced switching and tuning of the SC metamaterials.[11-13] The diverse behaviors of the transmission spectra varied with the temperature and the intense field indicate obviously that the heat and the nonlinearity produce different effects on the resonance of the SC metamaterial.

To elucidate this furthermore, we now evaluate the performance of the chosen SC metamaterial by an equivalent RLC circuit,[20] where the resonant transmission amplitude is dependent on the effective resistance $R$ (proportional to the ohmic losses of the film), and the resonance frequency $\omega_0$, which is strongly dependent on the effective capacitance $C$ and inductance $L$, i.e., $\omega_0=(LC)^{-1/2}$. Here, the total inductance $L$ includes geometry inductance $L_g$ and kinetic inductance $L_k$ at superconducting state as $L= L_g+ L_k$. For a superconducting film with a thickness of $d$, the surface impedance can be expressed by:[21]

$$Z_{s,eff} = R_{s,eff} + jX_{s,eff} = \sqrt{\frac{j\omega\mu_0}{\sigma(\omega)}} \coth(d\sqrt{j\omega\mu_0\sigma(\omega)}) \quad , \quad (1)$$

where $R_{s,eff}$ and $X_{s,eff}=\omega L_k$ are the effective surface resistance and reactance, respectively. $\mu_0$ is the magnetic susceptibility in vacuum.

With the measured complex conductivity, we can get the values of $X_{s,eff}$ and $R_{s,eff}$ as a function of incident terahertz field at 4.5 K (shown in Fig.4 (a)). The $R_{s,eff}$ represents a remarkable increase with increasing $E_{in}$, resulting in a notable decrease of the resonance dip. On the other hand, $X_{s,eff}$ ($L_k$) remains almost unchanged because of the balance between the increase of the density of normal-state electrons (enhance screening capability to $E_{in}$) and the decrease of the density of Cooper pairs (enhance penetration of $E_{in}$). As a consequence, the resonance frequency of our chosen SC metamaterials keeps almost unchanged as $L_g$ and $C$ are almost fixed with varying $E_{in}$. In addition, Figure 4(b) shows the surface resistance $R_{s,eff}$ and reactance $X_{s,eff}$ as a function of temperature at low terahertz field strength of $E_{in}=E_0/16$. Because of the large changes of $R_{s,eff}$ and $X_{s,eff}$ with the temperature, we thus find that the resonance dip was modulated significantly accompanying with an obvious redshift of the resonance frequency with increasing the temperature. So far, it is seen that although the modulations of the resonance in the chosen SC metamaterial is essentially associated with the changes of the complex conductivity of the integrated SC NbN components, the mechanisms are different by the temperature and the intense terahertz field. By the irradiation of the intense THz pulse, superconductivity of NbN film is rapidly switched with the duration of the monocycle THz pulse, and the nonlinear response time is from sub-picoseconds to more than 1 ns,[19] while the superconductivity of NbN film is steadily suppressed by the heating effect from temperature.

To summarize, we have devised, fabricated, and characterized a superconducting NbN metamaterial, which offers a highly efficient playground for a strongly enhanced nonlinearity and also provides a new and straightforward way for switching light in the terahertz regime. The switchable functionality is achieved due to the nonlinear response of the lumped superconducting NbN elements in the metamaterial structure, where as the incident terahertz field strength is varied, the conductivity of NbN will be changed and thus a rise or fall in transmission of the metamaterial would be obtained. It was also demonstrated that the unusual spectra response of the proposed SC metamaterials caused by the nonlinearities differs obviously from that caused by the heating effect. The presented nonlinear SC metamaterial exemplifies a highly functional approach with extreme sensitivity to the external stimuli to observe and exploit nonlinear phenomena in the terahertz regime, and strong and fast nonlinearities with far low losses are unique and useful to realize novel devices with integrated lumped nonlinear superconducting components.

This work is supported by the National Basic Research Program of China (Nos. 2011CBA00107, 2011CBA00202), Grant-in-Aid A222460430, Core to Core, JSPS, National High-Tech R&D Program of China under Grant No. 2011AA010204, the National Natural Science Foundation (Nos. 61071009, 61027008 and 61007034) , the Specialized Research Fund for Doctoral Program of Higher Education (20090091110040) and the Priority Academic Program Development of Jiangsu Higher Education Institutions (PAPD).

**Figure Captions**

**FIG. 1** (Color online) (a) Schematic diagram of the superconducting metamaterial structure with geometrical parameters: $g = t = 5$ µm, $w = 10$ µm, $a = 50$ µm, and $p = 60$ µm. (b) Microscopy image of the superconducting terahertz metamaterial sample, where E and H represent the electric field and magnetic field.

**FIG. 2** (Color online) (a) Measured amplitude transmission spectra at various incident terahertz field strengths at 4.5 K and (b) corresponding simulated spectra. (c) and (d) are the measured real part and imaginary part of the complex conductivity of NbN film having a same thickness of 50 nm at various incident terahertz field strengths.

**FIG. 3** (Color online) The amplitude transmission spectra of NbN metamaterial with various incident terahertz electric field at (a) 10 K and (b) 13 K, respectively. (c) the measured transmission spectra under various temperatures at low incident terahertz field of $E_0/16$.

**FIG. 4** (Color online) Effective surface reactance $X_{s,eff}$ and resistance $R_{s,eff}$ of the NbN film at resonance frequency of 0.45 THz (a) as a function of the incident terahertz field strength at 4.5 K and (b) as a function of the temperature with the low $E_{in}=E_0/16$.

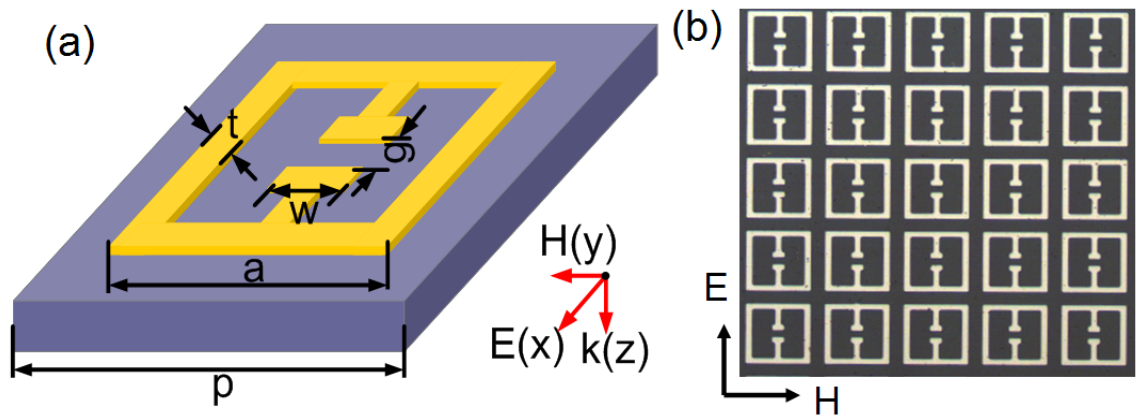

**FIG. 1.**
Zhang *et al.*

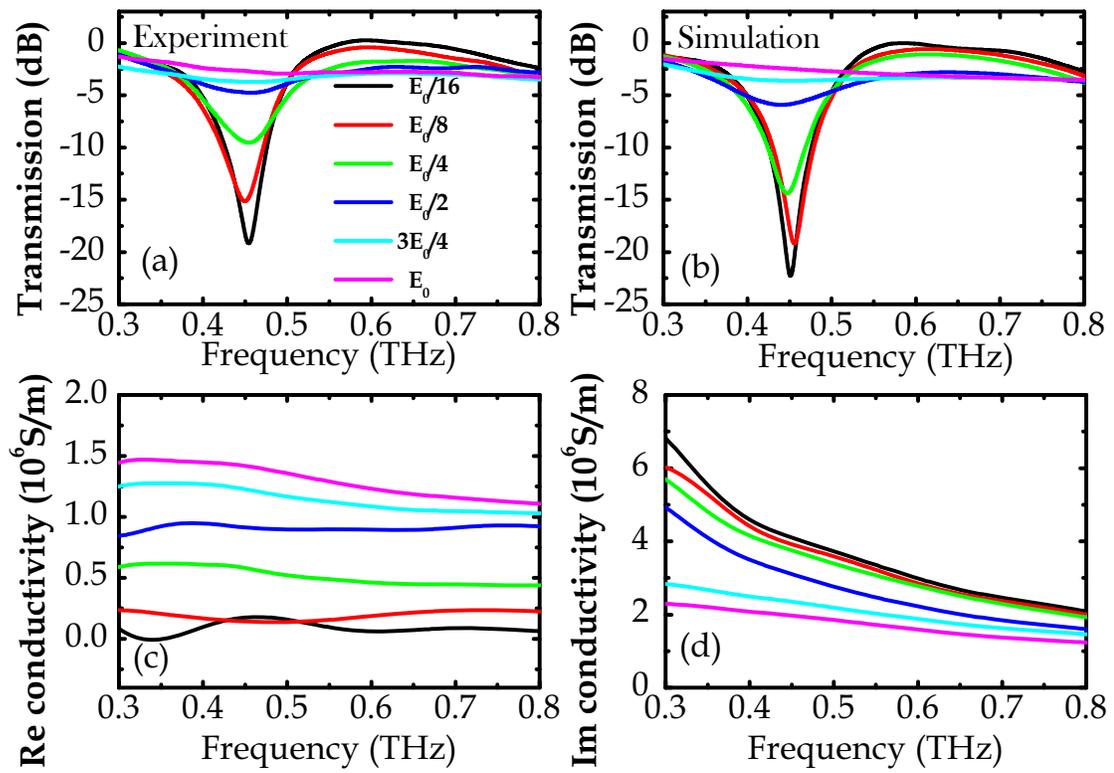

**FIG. 2.**
**Zhang** *et al.*

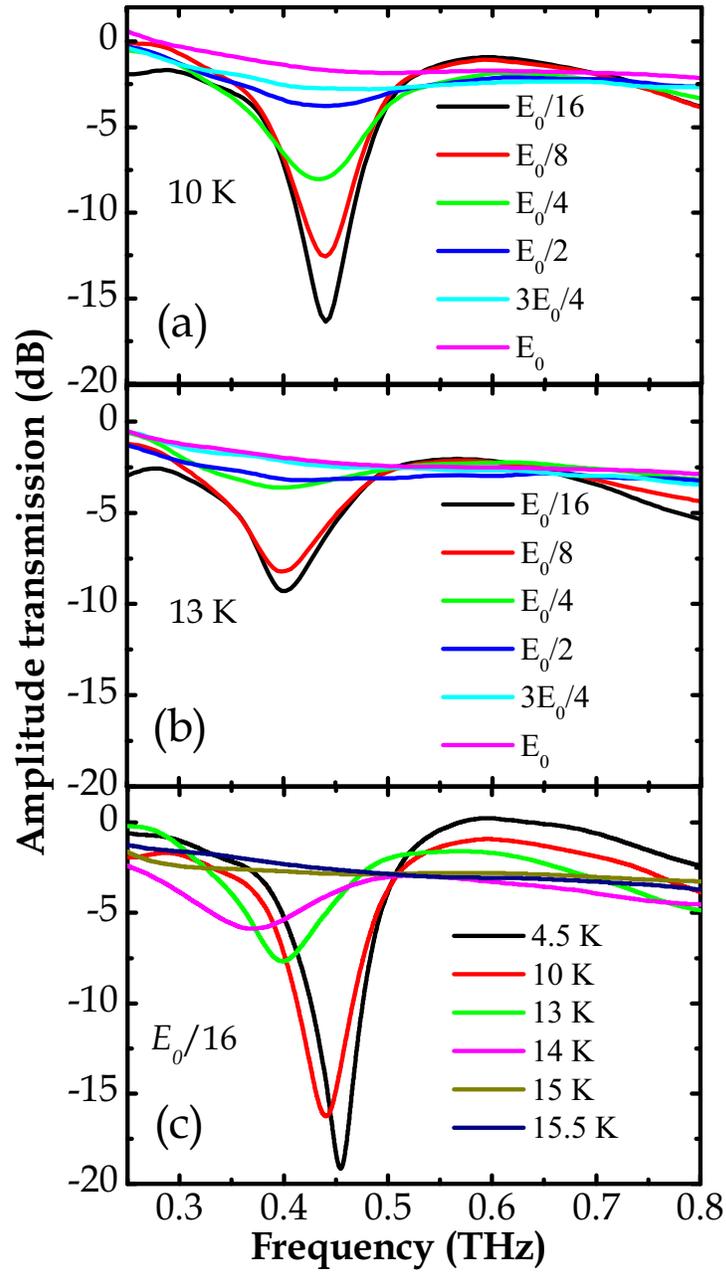

**FIG. 3.**
Zhang *et al.*

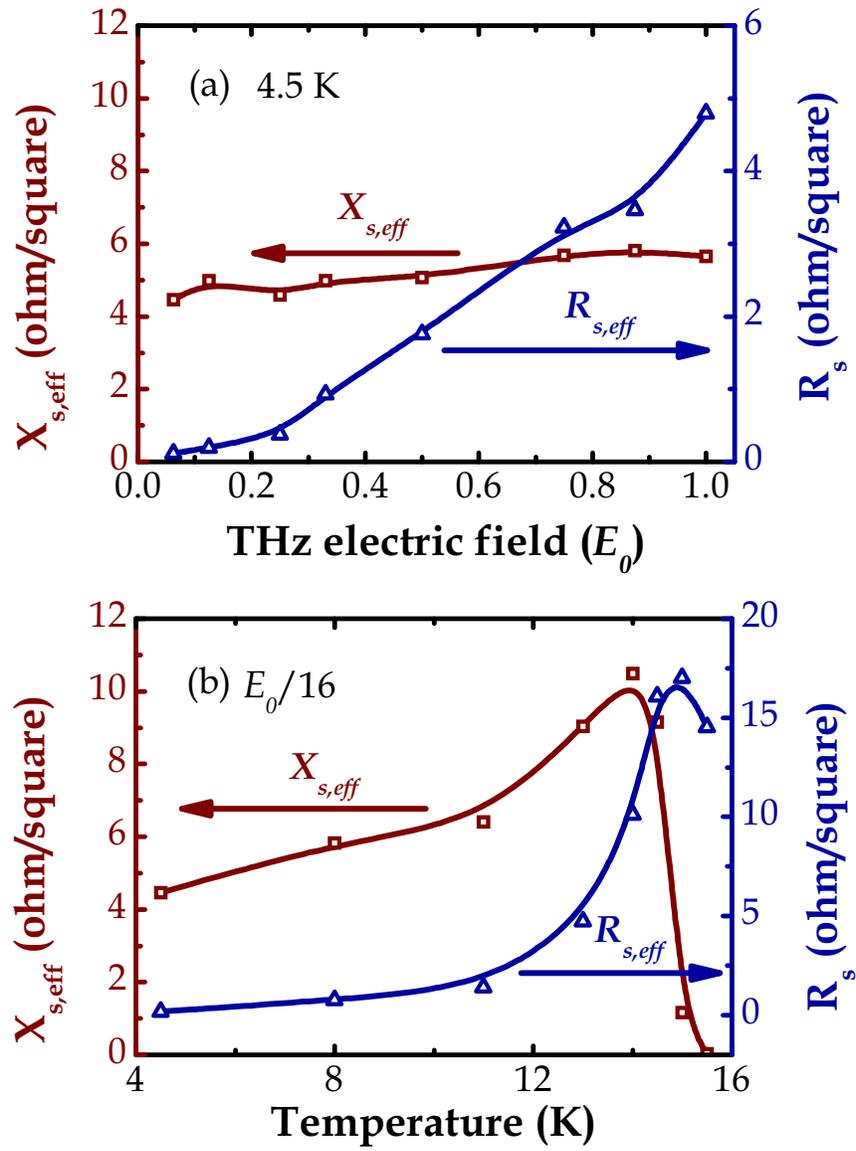

FIG. 4.
Zhang *et al.*